\documentclass{ws-ijmpa}

\begin{document}

\markboth{Kamal K. Seth}{Quarkonia \& Glueballs}

%
\catchline{}{}{}{}{}
%

\title{QUARKONIA \& PENTAQUARKS}

\author{Kamal K. Seth}

\address{Department of Physics and Astronomy, Northwestern University,\\
Evanston, IL 60208, USA\\
kseth@northwestern.edu}

\maketitle

\begin{abstract}
A brief review of the latest developments in the spectroscopy of heavy quarks is presented.  The current status of the recently `discovered' pentaquarks is also discussed.

\keywords{Quarkonia; Pentaquarks.}
\end{abstract}

\ccode{PACS numbers: 14.40Gx, 14.20Gk}

I am good at counting one, two.  Three is difficult for me.  So, I generally talk about mesons, and stay away from baryons.  When the organizers asked me to talk also about pentaquarks, that became a real challenge.  Since Frank Wilczek has told you all about the theoretical ideas behind pentaquarks, my job has become easier. I just have to tell you about experimental facts.

\section{Heavy Quarkonia}	

Light quark ($n\equiv u,d,s$) spectroscopy is very rich, and very tough.  The quarks are highly relativistic in the hadrons they make, the strong coupling constant $\alpha_s$ is very large ($\sim0.6$), and the $u,d,s$ quarks have such similar masses that nearly all $|n\bar{n}>$ mesons are mixtures of all three flavours.  This results in a very high density of overlapping states, difficult to disintangle, and even more difficult to understand.  In contrast, the charm ($c$) and beauty ($b$) quarks are heavy enough so that relativistic problems are not too serious ($\left<v^2/c^2\right>\approx 0.1-0.2$), $\alpha_s$ is not too large, ($\alpha_s\approx0.2-0.3$), and charmonium $|c\bar{c}>$ and bottomonium $|b\bar{b}>$ states are few and well resolved (see Fig. 1).  This makes the spectroscopy of $|c\bar{c}>$ and $|b\bar{b}>$ particularly useful for the study of Quantum Chromodynamics.

\subsection{Charmonium}

This is where all of it began with the 1974 discovery of $J/\psi$.  From 1974--1985 a lot of discovery physics in charmonium was done at SLAC, ORSAY and DESY, but precision was often lacking, except for the vector states which could be directly produced in $e^+e^-$ annihilation.  Width of triplet P-wave states could not be determined, and except for the ground state, no singlet S- and P-wave states could be successfully identified.  The region above the $D\bar{D}$ threshold essentially remained.

\begin{figure}[bt]
\centerline{\includegraphics[width=4.8in]{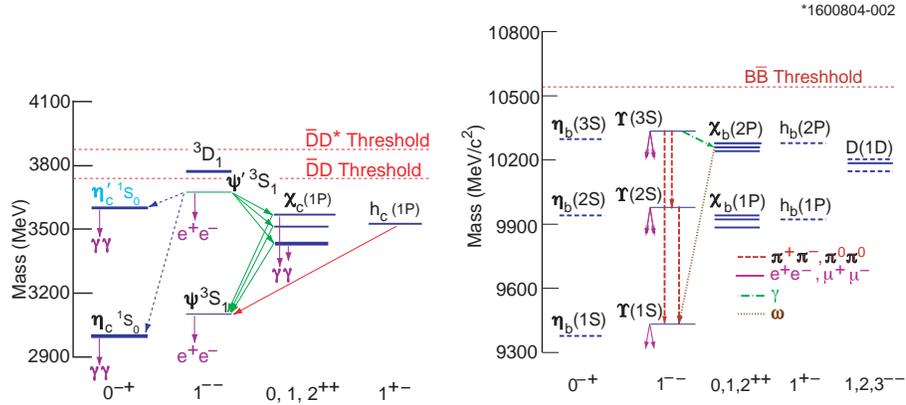}}
\vspace*{8pt}
\caption{Spectra of quarkonium states, (left) charmonium, (right) bottomonium.}
\end{figure}

During 1990--2000, the Fermilab experiments E760 and E835 exploited the ability of $p\bar{p}$ annihilations to make precision measurements of the masses and widths of $^3S_1$ ($J/\psi$, $\psi'$), and $1^3P$ ($\chi_0$, $\chi_1$, $\chi_2$) states, but were not so successful in making equally precise measurements of the singlet state $1^1S_0$ ($\eta_c$), and failed in identifying $2^1S_0$ ($\eta_c'$) and $1^1P_1$ ($h_c$).  The region above $D\bar{D}$ threshold remained \textit{terra incognita}.  For a review see Ref. 1.

During the 1990s, the BES detector at the Beijing Electron-Positron Collider (BEPC) made important contributions in charmonium spectroscopy, primarily by investing much greater luminosity ($\sim\times10$) then SLAC+ORSAY+DESY.  They also made some important excursions in the region above the $D\bar{D}$ threshold.

More recently, new players have emerged in the field.   The CLEO detector at the CESR accelerator at Cornell, the Belle detector at KEK, and the BaBar detector at PEP II at Stanford, are all beginning to produce extremely interesting results.  I am going to describe some of these below.  In somewhat more distant future (2008 --) we expect new accelerators, BEPC II and FAIR at GSI, to come online and provide further insight into the physics of this mass region.

\subsubsection{The Spin Singlet States and the Hyperfine Interaction}

The spin-indepedent $q\bar{q}$ interaction is well understood in terms of one-gluon exchange, and is very successfully modeled by a Coulombic 1/$r$ potential.  The spin dependence which follows from this is also accepted.  What is not understood is the the nature of the confinement part of the interaction, which is generally modeled by a scalar potential proportional to $r$.  A crucial test of the Lorenz nature of the confinement potential is provided by the measurement of hyperfine or spin-singlet/spin-triplet splittings.  A scalar potential does not contribute to the spin-spin or hyperfine interaction, whereas for a Coulombic potential it is a contact interaction.  As a consequence hyperfine splitting is predicted to be finite only for S-wave states, and to be zero for P-wave and higher L-states

\subsubsection{Hyperfine Splitting in S-wave Quarkonia}

No singlet states have so far been identified in bottomonium.  In charmonium, however, it has been established for a long time that $\Delta M(1S)_{hf}\equiv M(J/\psi, 1^3S_1)-M(\eta_c, 1^1S_0)=172\pm2$ MeV.  It is interesting to determine the size of the hyperfine splitting of $2S$ states, which sample the confinement region more deeply.  Long ago Crystal Ball\cite{cball} claimed the identification of $\eta_c'$ with $M(\eta_c')=3594\pm5$ MeV, leading to $\Delta M(2S)_{hf}=92\pm5$ MeV, which kind of made sense with $\Delta M(1S)_{hf}=172\pm2$ MeV.  Most potential model calculations tried to accomodate this `experimental' result , although it was not confirmed by any subsequent measurement, and was actually dropped by the PDG meson summary.

The seach for $\eta_c'$ has finally ended. Belle\cite{bellea} announced it first in two different decays of large samples of B-mesons.  CLEO\cite{cleoa} and BaBar\cite{babara} both have identified it in the two-photon fusion reaction, $e^+e^-\to(e^+e^-)\gamma\gamma,\gamma\gamma\to\eta_c'\to K_SK\pi$.  The CLEO measurement is shown in Fig. 2. The exciting part of these measurements is that $M(\eta_c')_{avg}=3637.4\pm4.4$ MeV, which is almost 50 MeV larger than the old Crystal Ball claim, and it leads to a surprisingly small hyperfine splitting, $\Delta M(2S)_{hf}=48.6\pm4.4$.  It is too early to say whether this can be explained in terms of channel mixing\cite{elq}, or unexpected contribution from the confinement potential.

\begin{figure}[tb]
\centerline{\includegraphics[width=2.25in]{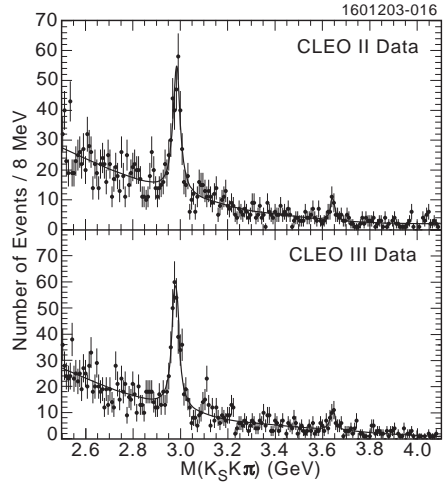}}
\vspace*{8pt}
\caption{CLEO discovery of $\eta_c'(2^1S_0)$ in two photon formation, and decay into $K_S K^\pm\pi^\mp$.}
\end{figure}

\subsubsection{Hyperfine Splitting in P-wave Quarkonia}

As mentioned already, hyperfine splitting is expected to be zero in all except S-wave states if the confinement potential is scalar, as is generally assumed.  Thus it is expected that $\Delta M(1P)_{hf} \equiv \left<M(1^3P_J)\right>-M(1^1P_1)=0$, except for higher order contributions of no more than an MeV or two.  Unfortunately, while $\left<M(1^3P_J)\right>=3525.31\pm0.07$\cite{cester}, the $h_c(1^1P_1)$ has not been firmly identified.  Let me however, give you a preview with the statement that both Fermilab E835 and CLEO are working on the search for $h_c$.  The E835 experiment is analyzing the reactions $p\bar{p}\to h_c\to\pi^0 J/\psi$ and $p\bar{p}\to h_c\to\gamma\eta_c$, and preliminary results are that while the first reaction does not have a signal for $h_c$ formation\cite{dave}, the second may have.  The CLEO team is analyzing $e^+e^-\to\psi'\to\pi^0 h_c,h_c\to\gamma\eta_c$ but has not presented any results so far [Note: Since the conference, CLEO has announced its preliminary results with $M(h_c)=3524.8\pm0.7$ MeV\cite{amiran} with the consequent $\Delta M(1P)_{hf}=0.6\pm0.6$ MeV. It appears that there is no significant departure from the simple expectation, $\Delta M(1P)_{hf}$=0].

\subsubsection{The $\rho-\pi$ Problem}

Since the widths for leptonic decays, as well as 3 gluon decays to light hadrons, of both $J/\psi$ and $\psi'$ depend on the wave functions at the origin, pQCD predicts the equality of the ratios of branching ratios
$$\frac{B(\psi'\to l^+l^-)}{B(J/\psi\to l^+l^-)} = (13\pm2)\% = \frac{B(\psi'\to LH)}{B(J/\psi\to LH)}.$$
This expectation has been extended to ratios of individual hadronic decays, and has led to many measurements by BES and CLEO to test it.  The results is that while the sums of all hadronic decays do seem to follow this expectation, and $\sum_i B(\psi'\to LH)_i/\sum_i B(J/\psi\to LH)_i = (17\pm3)\%$, individual decays show large departures from it, the ratio being as small as 0.2\% for $\rho\pi$ decays.  While many exotic theoretical suggestions have been made to explain these deviations, it appears that what we are witnessing is the failure of attempts to stretch pQCD beyond its limits of validity.

\subsubsection{Higher Vector States}

For a long time the parameters listed in the PDG compilation for the three vector states above the $D\bar{D}$ threshhold have been based on the R-parameter measurement by the DASP group\cite{dasp}, even though none of the other measurements of R agreed with it.  Recent measurements by the BES group\cite{besa} have finally allowed us\cite{sethb} to make a reliable determination of these parameters.  The result is that the total and leptonic widths of these states have changed by large amounts, e.g. $\Gamma(4039)=88\pm5$ MeV, instead of $52\pm10$ MeV.  Similar other new results are $\Gamma(4153)=107\pm8$ MeV, $\Gamma(4426)=119\pm15$ MeV.

\subsection{Bottomonium}

Despite the fact that the bottomonium $b\bar{b}$ system is certainly more amenable to pQCD, we know far less about bottomonium than we know about charmonium.  The $\eta_b$, ground state of bottomonium, has not been identified so far.  The vector states $\Upsilon(1S,2S,3S)$ and $4S,)$ are known but only one hadronic transition from these, $\Upsilon(nS)\to\Upsilon(n'S)\pi^+\pi^-\; (n'<n)$ has ever been observed.  Radiative transitions $\Upsilon(nS)\to\gamma\chi_b(n'^3P)$ states have been observed.  No D-states, which are expected to be bound (see Fig. 1) have been observed.  No hadronic transition from any $\chi_b$ states has ever been observed.  Recently, CLEO had made small gains in both the above problems. The $1^3D_2$ state has been successfully observed in 4-photon cascade $\Upsilon(3S)\to\gamma_1(2P)\to\gamma_1\gamma_2(1D)\to\gamma_1\gamma_2\gamma_3(1P)\to\gamma_1\gamma_2\gamma_3\gamma_4\Upsilon(1S),\Upsilon(1S)\to l^+l^-$.  The mass $M(1^3D_2)=10,161.1\pm0.6\pm1.6$ MeV\cite{cleob}.  In another measurement, $\Upsilon(3S)\to\gamma\chi_b(2P),\chi_b(1,2)\to\omega\Upsilon(1S)$ has also been observed\cite{cleoc}.

\section{Exotics}

I was going to talk about the classical exotics of QCD, the glueballs and hybrids, but because of the request of the conference organizers to talk about pentaquarks, I will simply refer you to my last review of glueballs and hybrids\cite{sethc}.  To summarize, no consensus candidates for scalar or tensor glueballs have emerged so far.  Candidates for $|q\bar{q}g>$ hybrids with exotic $J^{PC}=1^{-+}$ have indeed been claimed amid plenty of controversy.

Having been released from glueballs and hybrids, I can indulge in another class of exotics.  These are the unexpected, uninvited, and therefore the exotic hadrons which have recently shown up in several hadron spectroscopy experiments.

The first of these was the discovery of narrow ($\Gamma<7$ MeV) $D_{sJ}$ resonances with $M(D^{*+}_s,J^{P}=0^+)=2317$ MeV, and $M(D^{*+}_s,J^{P}=1^+)=2462$ MeV by BaBar\cite{babarb} and CLEO\cite{cleod}.  These states were expected at higher masses and therefore with large widths.  Instead, they show up temptingly just below $DK$ and $D^*K$ thresholds, giving encouragement to molecular enthusiasts.

The second exotic is the discovery by Belle\cite{belleb} of a narrow resonance in $B$ decays, which was quickly confirmed by CDF.  This resonance, dubbed X(3872), has a mass of $3872\pm1$ MeV, a width $<2.5$ MeV, and decays (it appears almost exclusively) to $\pi^+\pi^- J/\psi$.  Again, since its mass, width, and decay make it difficult to fit it into the charmonium spectrum, and since $M(D^0)+M(\bar{D}^{*0})=3872$ MeV, it has provided more fodder to $|D^0\bar{D}^{*0}>$ molecule enthusiasts.  At CLEO we have searched for this state in untagged two photon fusion (therefore $J^{PC}=0^{\pm,+},2^{\pm,+},...$) and in ISR (initial state radiation) mediated production, and have established quite stringent upper limits on its population in either production mode\cite{cleoe}.

I finally come to the hotter than hot topic of...

\section{Pentaquarks}

The pentaquark story starts with the annoucement by LEPS at Osaka\cite{nakano} that there was a significant enhancement in the missing mass spectrum for $\gamma K^-$ in the reaction $\gamma n \to K^+K^- n$ with photons incident on a plastic scintillator (CH) target.  The missing mass spectrum was interpreted to indicate a $|K^+n>$ state dubbed $\Theta^+$, with mass $M(\Theta^+)=1.54\pm0.01$ GeV, and width $\Gamma(\Theta^+)<25$ MeV, with statistical significance of $4.6\sigma$.  If true, the state had strangeness $S=+1$, and had to have at least five quarks.  The pentaquark was born!  In quick succession, CLAS claimed confirmation in photons incident on deuterium\cite{clasa}, and hydrogen\cite{clasb}, SAPHIR in $\gamma+p$\cite{saphir}, ZEUS\cite{zeus} in $e^+p$ and $e^-p$ inelastic scattering, HERMES\cite{hermes} in $e^+d$ inelastic scattering, YEREVAN in $p$+propane\cite{yerevan}, DIANA\cite{diana} in $K^+$+Xenon, SVD in $p$+Si\cite{svd}, and COSY in $p+p$\cite{cosy}.  Neutrinos were also not left behind, and ITEP\cite{itep} claimed $\Theta^+$ in $\nu+$H$_2$,D$_2$,Ne data from CERN and FNAL.  In my memory, never before has such a stampede been witnessed!

The theoretical model for $\Theta^+$ which was in vogue is the antidecuplet model of Jaffe and Wilczek\cite{jw}.  According to this model, there should be $(S=0)\;N^*$, $(S=-1)\;\Sigma$, and $(S=-2)\Xi$ cascade pentaquarks also.  Sure enough, NA49\cite{na49} announced the observation of $\Xi(1862)$ as the $S=-2$ pentaquark in the reaction $p+p\to X+(\Xi^-\pi^-)$.  Going one step further, H1\cite{h1} claimed the observation of a charmed pentaquark $\Theta_c(3099)$ in the reaction $e+p\to (D^*p)$.

One would think that with so many positive observations there is no doubt that pentaquarks exist.  Unfortunately, this is not so.  There are two reasons.  The first is illustrated in my Table I.  The fact is that despite the claimed significance levels up to $7.8\sigma$, a conservative uniform determination of significance, $\sigma=S/\sqrt{S+2B}$, where $S$ and $B$ are signal and background counts respectively, leads to the fact that none of the significance levels rises to the level of $5\sigma$, the criterion used by Physical Review Letters for a claim of observation.

\begin{table}[bt]
\begin{center}
\begin{tabular}{l|l|l|l|l|l|l}
\multicolumn{7}{c}{\rule{0cm}{.0cm}}\\
\hline
\hline
Mass & Width & $N$ & Signif. & Uniform  & Reaction & Experiment \\
(MeV) & (MeV) & & Claimed &  signif.  & $A + B \rightarrow X + \Theta$ & \\
\hline
$\Theta^{+}(1540)$  & & & & & & \\
$1540 \pm 10 \pm 5$ & $< 25$ & 19 & 4.6 $\sigma$ &  $\sim 2.7\sigma$ & $\gamma +$ C $\rightarrow X + (n K^{+})$ & LEPS \\
$1542 \pm 2 \pm 5$ & $< 21$ & 43 & 5.2 $\sigma$ &  $\sim 3.5\sigma$  & $\gamma + d$ $\rightarrow X + (nK^{+})$ & CLAS \\ 
$1540 \pm 4 \pm 3$ & $< 25$ & 63 & 4.8 $\sigma$ &  $\sim 4.3\sigma$  & $\gamma + p$ $\rightarrow X + (nK^{+})$ & SAPHIR \\ 
$1555 \pm 1 \pm 10$ & $< 26$ & 41 & 7.8 $\sigma$ &  $\sim 4.0\sigma$ & $\gamma + p$ $\rightarrow X + (nK^{+})$ & CLAS \\ 
 & & & & & & \\
$1539 \pm 2 \pm 2$ & $< 9$ & 29 & 4.4 $\sigma$ & $\sim 3.0\sigma$  & $K^{+} +$ Xe $\rightarrow X + (pK^0_s)$ & DIANA \\ 
$1533 \pm 5 \pm 3$ & $< 20$ & 27 & 6.7 $\sigma$ & $\sim 4.0\sigma$  & $\nu +$ Ne $\rightarrow X + (pK^0_s)$ & ITEP \\ 
$1528 \pm 4$ & $< 19$ & 60 & 5.8 $\sigma$ & $\sim 4.0\sigma$  & $\gamma^{*} + d$ $\rightarrow X + (pK^0_s)$ & HERMES \\ 
$1526 \pm 3 \pm 3$ & $< 24$ & 50 & 5.6 $\sigma$ & $\sim 3.5\sigma$ & $p +$ Si $\rightarrow X + (pK^0_s)$ & SVD-2 \\ 
$1530 \pm 5$ & $< 18$ & 50 & 3.7 $\sigma$ & $\sim 3.7\sigma$ & $p + p$ $\rightarrow X + (pK_s^0)$ & COSY \\ 
$1545 \pm 12$ & $< 40$ & 100 & 5.5 $\sigma$ &  $\sim 4.0\sigma$ & $p +$ prop $\rightarrow X + (pK^0_s)$ & YEREVAN \\ 
$1522 \pm 2 \pm 3$ & $< 6$ & 221 & 4.6 $\sigma$ &  $\sim 3.6\sigma$  & $\gamma^{*} + p$ $\rightarrow X + (p^{\pm}K_s)$ & ZEUS\\ 
\hline
$\Xi(1862)$ & & & & & & \\
 $(S + -2)$  & & & & & & \\
$1862 \pm 2$ & $< 21$ & 65 & 5.8 $\sigma$ &  $\sim 4.7\sigma$ & $p + p$ $\rightarrow X + (\Xi^{-}\pi^{-})$ & NA49 \\ 
\hline
$\Theta(3099)$ & & & & & & \\
$(C = -1)$ & & & & & & \\
$3099 \pm 3 \pm 5$ & $< 35$ & 51 & 5.4 $\sigma$ & $\sim 4.2\sigma$ & $e + p$ $\rightarrow X + (D^{*}p)$ & HERA \\ 
\hline
\hline
\end{tabular}
\end{center}
\caption{Summary of claimed observations of pentaquarks.  The uniform significance levels have been calculated as $\sigma=S/\sqrt{S+B}$ with estimated signal $(S)$ and background $(B)$ counts when they are not explicitly given.}
\end{table}

The second reason is much more important.  Since pentaquarks are so novel and exciting, many more experiments have attempted to find them, but failed.  Those which have tried, but failed to find $\Theta^+$ include BES\cite{besb}, FNAL E690\cite{e690}, FNAL E871 (HyperCP)\cite{e871}, CDF\cite{cdf}, BaBar\cite{babarc}, ALEPH\cite{aleph}, DELPHI\cite{delphi}, PHENIX $(\bar{\Theta})$\cite{phenix}, and HERA-B\cite{herab}.  Similarly, CDF\cite{cdf}, BaBar\cite{babarc}, ALEPH\cite{aleph}, HERA-B\cite{herab}, and ZEUS\cite{zeusb} find no evidence for $\Xi(1860)$.  CDF\cite{cdf}, ALEPH\cite{aleph}, ZEUS\cite{zeus}, and FNAL E831 FOCUS\cite{focus} also find no evidence for $\Theta_c$.  It is not looking very good for the survival of any of the pentaquarks!

Finally, I have to confess to my personal skepticism about pentaquarks.  Nearly two decades ago, there was a similar stampede for the existence of dibaryons, and for several years I made valiant searches for them.  While claims for nearly 40 dibaryons with masses between 1900 and 2250 MeV were made, not a single one survived high resolution, high statistics measurements\cite{sethd}.

This research was supported by the U. S. Department of Energy.

Note: The references listed below include several which were not published at the time of the conference, but have become available since then.

\end{document}